\documentstyle [12pt,epsf,epsfig] {article}
\begin{document}
\bibliographystyle{unsrt}

\pagestyle{empty}               
	
\rightline{\vbox{
	\halign{&#\hfil\cr
	&BNL- \cr}}}

\rightline{\vbox{
	\halign{&#\hfil\cr
	&September 1996\cr}}}
\vskip 1in
\begin{center}
{\Large\bf
{CNI Polarimetry and the Hadronic Spin Dependence of pp Scattering }}
\vskip .5in
\normalsize
T.L.\ Trueman \footnote{This manuscript has been authored
under contract number DE-AC02-76CH00016 with the U.S. Department
of Energy.  Accordingly, the
U.S. Government retains a non-exclusive, royalty-free license to
publish or reproduce the published form of this contribution, or
allow others to do so, for U.S. Government purposes.}\\
{\sl Physics Department, Brookhaven National 
Laboratory, Upton, NY 11973}
\end{center}
\vskip 1.5in
\begin{abstract}
Methods for limiting the size of hadronic spin-flip in the
Coulomb-Nuclear Interference region are critically assessed. This work was
presented at the High Energy Polarimetry Workshop in Amsterdam, Sept.9,
1996 and the RHIC Spin Collaboration meeting in Marseille, Sept. 17, 1996.
\end{abstract}
\vfill \eject \pagestyle{plain}
\setcounter{page}{1}

The interference between the Coulomb spin-flip amplitude $\phi_5$
and the hadronic non-flip amplitude $\phi_1 + \phi_3$ at small $t$
contributes a calculable amount to $A_N$, an amount that
is small but large enough to be used as a practical polarimeter for
colliding proton beams \cite{AnnArbor}. However, the corresponding
interference between the {\it hadronic} spin-flip amplitude and the
{\it Coulomb} non-flip amplitude contributes a piece to $A_N$ that
has exactly the same shape for small $t$ as does the CNI piece.
Thus in order for CNI to be useful as a polarimeter it is
necessary that a suitably small bound be known for the hadronic
spin-flip amplitude. This has been discussed in an earlier note
\cite{TLT} where all of the standard notation is recapitulated.
 Here we will extend that discussion in some small ways.
 
The first question is, can the size of $\phi_5$ be limited by
measurements taken in the same experiment that is used to measure
$A_N$, in particular the pp2pp experiment at RHIC \cite{pp2pp}? To discuss
this, we parametrize the hadronic spin-flip amplitude in terms of
the non-flip amplitude as
\begin{equation}
\phi_5^h = \tau \sqrt{-t/m^2} (\phi_1^h +\phi_3^h)/2
\end{equation}
In general $\tau$ is a complex function of $s$, the total energy
squared, but we will assume that it is independent of $t$ at least
for $|t|<0.05  {\rm GeV}^2$. It may be constant for a larger region,
but it seems unlikely that this will be so for $|t|$ much greater than
$0.1 {\rm GeV}^2$. In this region for high energy, the analyzing power is
given by
\begin{equation}
A_N \frac{d\sigma}{dt} = \frac{\alpha \sigma_{tot}e^{bt/2}}{2 m
\sqrt{-t}}
\{(\mu - 1) - 2 {\rm Re}(\tau) - 2 \rho {\rm Im}(\tau) \} + 2 {\rm Im}(\tau)
\frac{\sqrt{-t}}{m}\left(\frac{d\sigma}{dt}\right)_{\rm hadronic}
\end{equation}
From this expression we see two important features. The first is
that ${\rm Re}(\tau)$ just shifts the CNI curve up or down; it does not
modify the shape at all. Thus it is impossible to infer a bound on
${\rm Re}(\tau)$ from the shape of $A_N$ in the CNI region.If the best
limit we have on ${\rm Re}(\tau)$ is 
\begin{equation}
|{\rm Re}(\tau)| \leq \tau^*
\end{equation}
then the precision with which the polarization $P$ can be measured
is limited by
\begin{equation}
\frac{\Delta P}{P} \geq \frac{2 \tau^*}{(\mu -1)}.
\end{equation}
In particular, a 5\% measurement of $P$ requires that $\tau^* <0.05$.

Second, the shape of the curve is evidently quite sensitive to
${\rm Im}(\tau)$ because it leads to purely hadronic spin-flip in $A_N$.
However, this contribution is not enhanced at small $|t|$ and so the
CNI peak is not sensitive to it. (See Fig.1.)
\begin{figure}
\centerline{\epsfbox{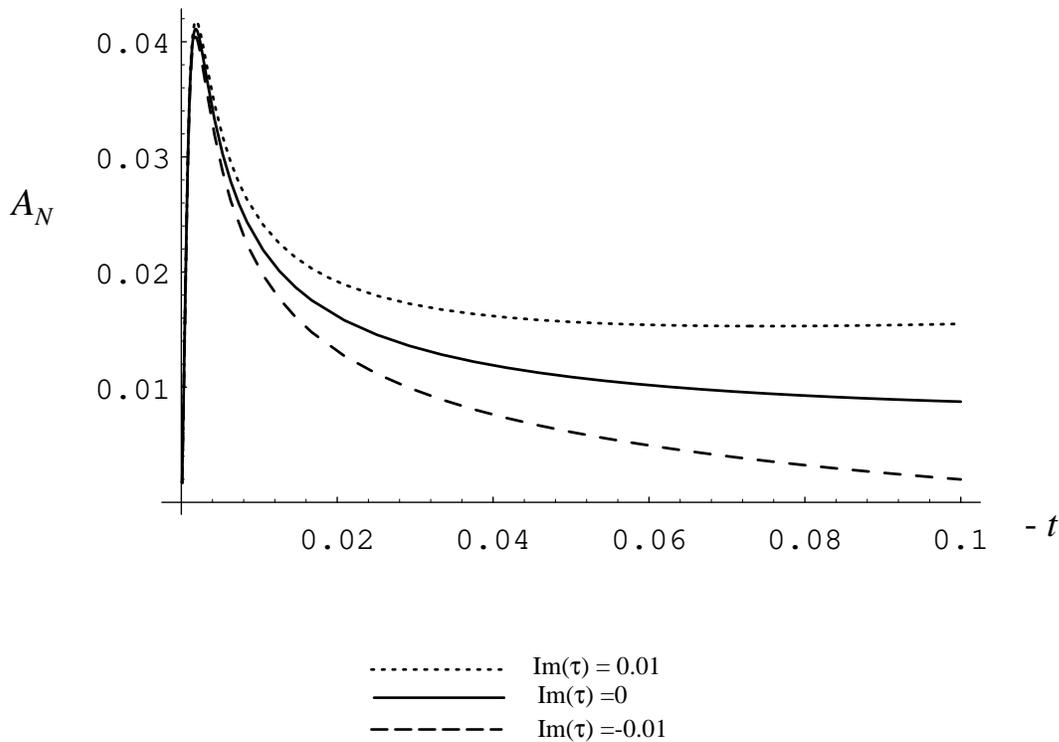}}
\medskip
\caption{$A_N$ at 500 {\rm GeV} cm energy for ${\rm Re}(\tau)=0$ and
${\rm Im}(\tau) = -0.01,0,+0.01$}
\end{figure}
Since {\it a priori} the phase of $\tau$ is
unknown, this sensitivity is not useful for constraining
the hadronic contribution to $A_N$ in the CNI region. Krisch and
Troshin \cite{KT} have argued that the phase of $\tau$ should not be
too small. In that case, one can use data at moderate values of $t$,
$0.1<-t<0.5$, to bound ${\rm Im}(\tau)$ and hence infer a bound on
${\rm Re}(\tau)$. On the contrary, if the elastic scattering is dominated
by the exchange of $C=+1$ in the $t$-channel, as it would be if
dominated by Pomeron and multiple-Pomeron exchange, crossing
relations for small $t$ and large $s$ imply that asymptotically the
amplitude is pure imaginary. This is true for the spin-flip just as
for the non-flip amplitudes \cite{Eden,TW}. If $|\phi_1|$ 
and $|\phi_5|$ have exactly the same asymptotic
behaviour, then ${\rm Im}(\tau) = 0$. One also learns from this
argument that ${\rm Re}(\phi_5)/{\rm Im}(\phi_5) \propto 1/\ln s$. One
might reasonably estimate, by analogy with the non-flip
amplitude, that this is of order $\rho$; i.e. of order 0.1 and slowly
falling with energy.
If there is significant odderon contribution, having
the same asymptotic behaviour but with $C=-1$ this restriction on the
phase is no longer true. Indeed, the measurement of this phase would
be another way of investigating the presence of the odderon in
elastic pp scattering.

We have just seen that the
{\em shape} cannot be used to limit $\tau^*$. Examination of the
standard reference of Buttimore, Gotsman and Leader \cite{BGL} shows
that the spin-flip amplitude
$\phi_5$ contributes to three other measurable quantities (assuming
always that final polarizations are not measured): the differential
cross section, the two-spin asymmetry $A_{NN}$ and the two-spin
asymmetry
$A_{SL}$. In none of these is its contribution enhanced by
interference with the Coulomb amplitude, and so one expects in
advance that it will be difficult to get an adequate constraint from
these measurements.

Buttimore \cite{Buttimore} has worked out a constraint coming from
the differential cross section which we can write in the form
\begin{equation}
|\tau| < \sqrt{\frac{bm^2}{2} \left(\frac{16\pi b
\sigma_{el}}{(1+\rho^2)\sigma_{tot}^2}-1\right)}.
\end{equation}
Because $b m^2 > 10$ even a 10\% measurement of $\Delta P/P$ would
require that the combination in the parenthesis be known to better
than three parts in $10^3$. A 1\% measurement of that combination 
would lead to a bound of  $|\tau| < 0.24$ and so limit
the precision of $\Delta P/P$ to about 27\%. The use of this bound
requires a further analysis of how the quantities $b, \sigma_{el},$ and
$\sigma_{tot}$ are determined. For example, if $\sigma_{tot}$ is
determined by extrapolation of $d\sigma/dt$ accompanied by the usual
assumption of spin-independence, viz.$\phi_1 = \phi_3$ and $\phi_2 = 0$
additional uncertainties are introduced. See the talk by Andr\'e Martin at
the Marseille conference \cite{AM}.

Rather than pursue this, let us consider the differential cross
section directly. Continuing with the assumptions of Eq.(1), we have 
\begin{equation}
\frac{d\sigma}{dt} = \frac{1}{16 \pi} (1 + \rho ^2) \sigma _{tot}^2
(1 - 2 |\tau|^2 \frac{t}{m^2}) e^{bt},
\end{equation}
where $b$ is the slope of the diffraction peak in the small $t$
region, say $4 \times 10^{-4} \leq |t| \leq 0.2 {\rm GeV}^2$ as planned
in the pp2pp experiment. This form yields an effective slope $b'$,
where
\begin{equation}
b' = b - 2 \frac{|\tau|^2}{m^2} + 2 \frac{|\tau|^4 t}{m^4} + O(t^2).
\end{equation}
It is essential to limit the third term because the second term
just shifts $b$. For statistical errors of the magnitude which are
the goal of pp2pp, about $10^{-3}$, this yields at best a useless
bound of $\tau^* <0.6$, even disregarding other possible sources of $t$
dependence of $b$. Because
$\tau$ enters to the fourth power, it seems hopeless to pursue this line.

The situation with $A_{NN}$ is more problematic because, in addition
to $\phi_5$ it also depends on the unknown combination
\begin{equation}
{\rm Re}( \phi_1^* \phi_2 - \phi_3^* \phi_4).
\end{equation}
This piece is very likely to prevent a useful bound from being
obtained here. If we disregard the purely hadronic contribution to this
piece  we find that
\begin{equation}
A_{NN}\frac{d\sigma}{dt}=
\sigma_{tot} \exp ^{bt/2}
\frac{\alpha}{4m^2} (\mu-1)\{(\mu -1)\rho -4(\rho {\rm Re}(\tau) -
{\rm Im}(\tau))\} -2\frac{t}{m^2} |\tau|^2
\left(\frac{d\sigma}{dt}\right)_{\rm hadronic}
\end{equation}
If the errors on the double spin asymmetry are of 
order $10^{-3}$ for $t$ between 0.002 and 0.05 the best
bound this gives for $\tau^*$ is about 0.15. Unfortunately, rather
small double flip amplitudes $\phi_2$ and $\phi_4$ can cancel this
small asymmetry and destroy even this weak bound. This is illustrated in
Fig.2. Here we have assumed that
\begin{eqnarray}
\phi_1 & = & \phi_3 \\
\phi_2 & = & -\phi_4,
\end{eqnarray}
and introduced the natural parametrization
\begin{equation}
\phi_2 = -\delta \frac{t}{m^2} \phi_1 ,
\end{equation}
\begin{figure}
\centerline{\epsfbox{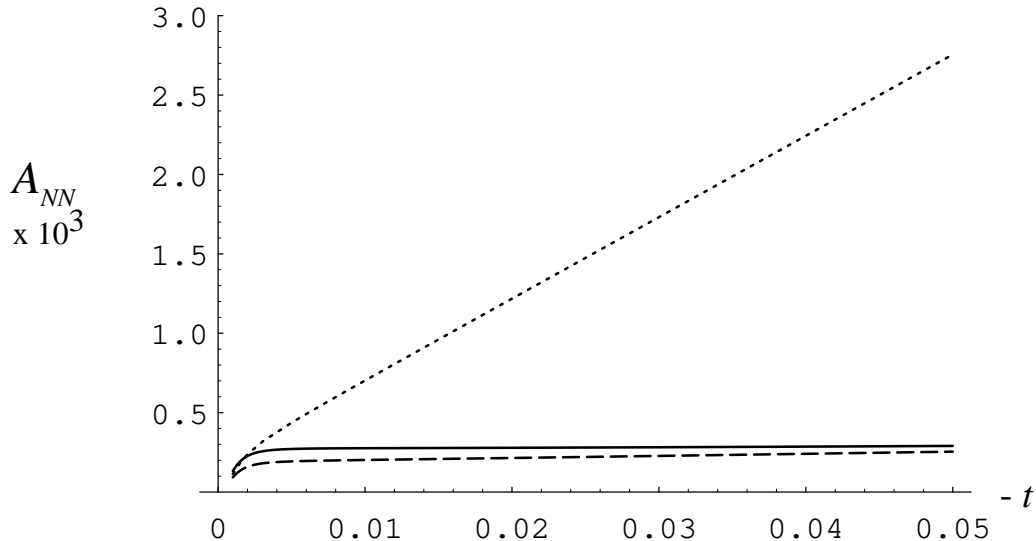}}
\medskip
\caption{$A_{NN}$ at 500 {\rm GeV} cm energy for $\tau = \delta = 0$
(solid), $\tau = 0.15$, $\delta = 0$(dotted) and $\tau = 0.15$,
$\delta = -0.022$ (dashed).}
\end{figure}
The amplitudes $\phi_2$ and $\phi_4$ are nearly two orders of
magnitude smaller than $\phi_5$ in this region, but essentially cancel its
effect, because these
enter $A_{NN}$ by interfering with the principal amplitudes $\phi_1$
and $\phi_3$.

(It is interesting to explore the possibility of using the fact that
the single spin asymmetry is proportional to $P$ while the double
spin asymmetry is proportional to $P^2$. Perhaps if one could get
the errors down this could be used to extract $P$? We continue using the
parametrization just introduced. Then $A_{NN}$ is the only two-spin
asymmetry which depends on $\phi_5$; in particular, $A_{SL}=0$,
independent of $\phi_5$. 
\cite{BGL}.
Again, it is very easy to find values of $\tau$ and $\delta$
such that $A_{NN}$ has the same shape as it would have for these
parameters set to zero, and furthermore it is shifted in magnitude
by the square of the factor that $A_N$ is ; 
i.e if we write 
\begin{equation}
P_{apparent}(\tau) = P_{true} \frac{A_N(\tau \neq 0)}{A_N(\tau = 0)}
\end{equation}
then we find over the CNI region that
\begin{equation}
A_{NN}(\tau =0.1,\delta=-0.01) \approx A_{NN}(0,0)
\left(\frac{P_{apparent}(0.1)}{P_{true}} \right)^2
\end{equation}
and so if in fact $\tau=0.1$ and $\delta = -.01$ while we assume
that $\tau = \delta = 0$ we would infer consistent but erroneous
values of the polarization from the two different measurements.
(See Fig.3.))
\begin{figure}
\centerline{\epsfbox{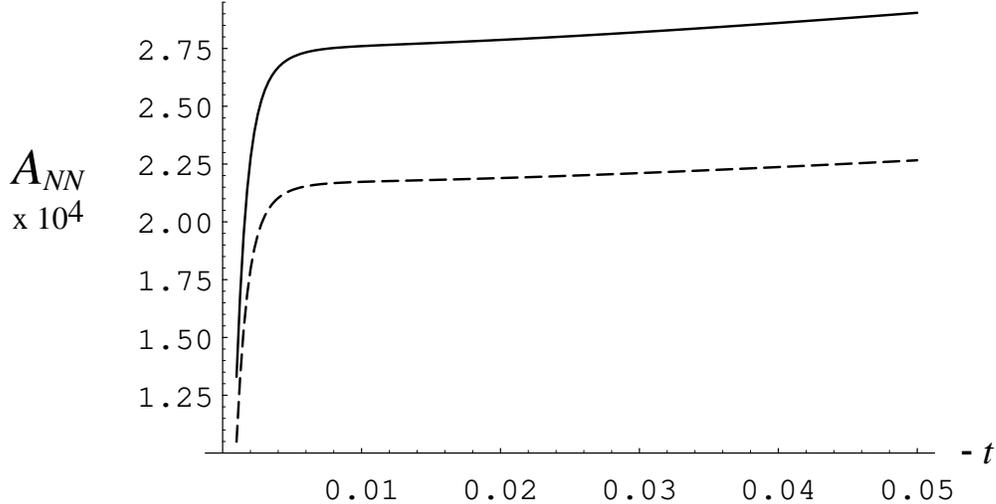}}
\medskip
\caption{$A_{NN}$ at 500 {\rm GeV} cm energy comparing $\tau = \delta =
0$ (solid) with $\tau = 0.1, \delta = -0.01$ (dashed)}
\end{figure}

I conclude from this analysis that it will not be possible to
constrain the size of the hadronic spin-flip amplitude using only
data from the pp2pp experiment. We then ask if we can get useful
constraints from other experiments.

The natural thing to do is to look at the size of $A_N$ as
determined from lower energy experiments at Fermilab and at CERN,
in the 1970's. One can then try to extrapolate these
measurements to higher energy by fitting their energy dependence.

A glance at the data indicates that $A_N$ is falling very fast with
energy and so it is tempting to believe that the hadronic spin-flip
amplitude will fall off to negligible levels by the time RHIC energy
is reached. In order to test this quantititatively we have taken a
collection of data from various experiments at different energy and
all for $t = -0.15 {\rm GeV}^2$ (or interpolated from nearby values), the
smallest
$|t|$ for which there is sufficient data to do this \cite{data}.
We have tried a fit suggested by Regge poles, namely $a +
b/\sqrt{p_L} + c/p_L$, where $p_L$ denotes the lab momentum for
these fixed target experiments. This is shown in Fig.4.
\begin{figure}
\centerline{\epsfbox{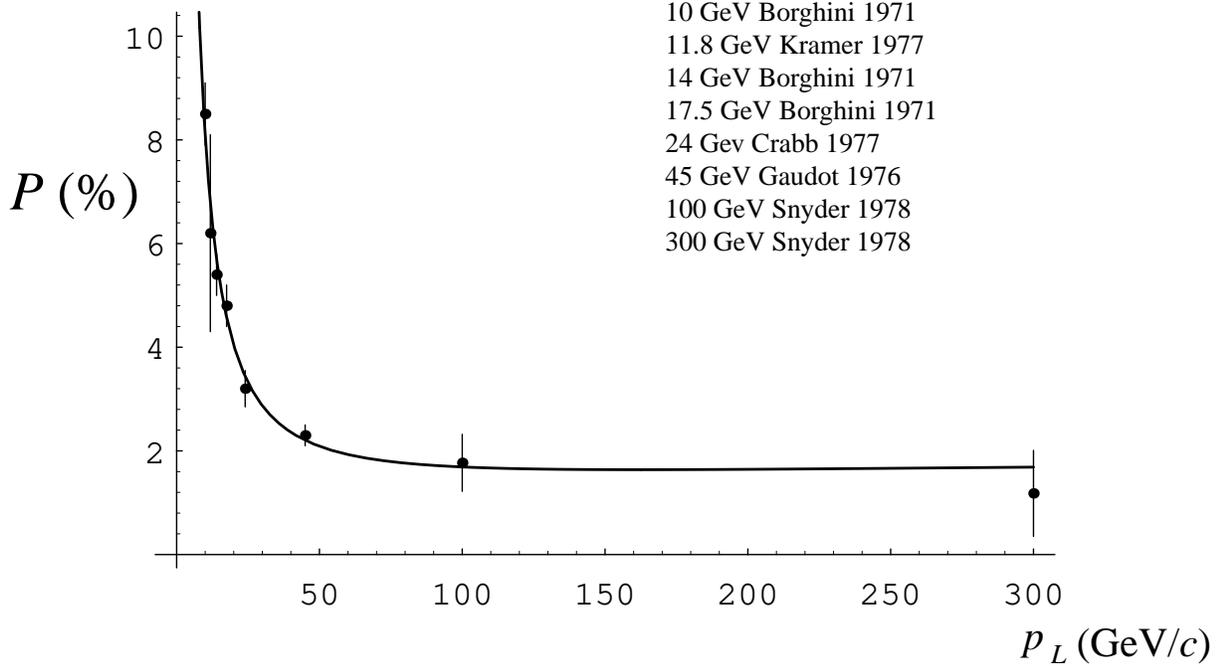}}
\medskip
\caption{Energy dependence of $A_N$ at $t = -0.15 {\rm GeV}^2$. The best
fit gives the asymptotic value 2.3 $\pm $ 1.2. The CNI value at 300
{\rm GeV} is 1.1}
\end{figure}
The $\chi^2$ is quite good. The relevant result is that 
\begin{equation}
a = .023 \pm .012
\end{equation}
which will be the extrapolated value of $A_N$ to very high energy.
This is not very well determined; it is consistent with pure CNI which is
approximately equal to .01 at this $t$-value. It is also consistent with
${\rm Re}(\tau) = -1$ ! Hence it can certainly not be used to limit the magnitude
of
${\rm Re}(\tau)$ to any useful value, even if we are willing to use the CNI
parametrization to values of $|t|$ this large. (I have also done a similar
fit at
$t = -0.3 {\rm GeV}^2$ but, while there is more data available there, the
scatter from experiment to experiment is quite large and the
$\chi^2$ is not good. The fit obtained is consistent with pure CNI.) A
fit with a pure $1/p_L$ behaviour, the type suggested by the kinematic
factor $\sin \theta /2$ in $\phi_5$, has a much worse $\chi^2$.

Experiment 704 at Fermilab \cite{704} is potentially the most
relevant experiment for our question. It measures $A_N$ by
elastically scattering protons of known polarization in the
CNI region. The errors are unfortunately rather large, but even so
this experiment probably gives us the best bound we have on
${\rm Re}(\tau)$. I have fit the data up to $|t|<.05{\rm GeV}^2$ with
the simple form that I have used throughout, assuming ${\rm Im}(\tau)=0$
and find the
best fit is within 1\% of the pure CNI curve. The error is about
15\%. See Fig.5. 
\begin{figure}
\centerline{\epsfbox{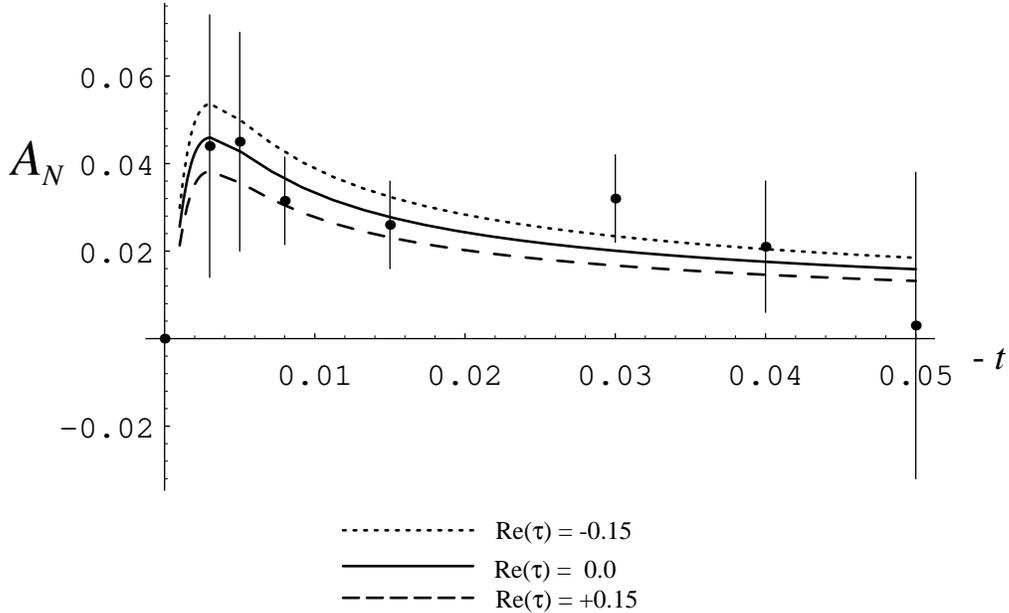}}
\medskip
\caption{Fit to E704  data with pure CNI and CNI with $\tau =
\pm 0.15$}
\end{figure}
If, however, we allow a non-zero ${\rm Im}(\tau)$ the best fit gives ${\rm Re}(\tau)=
0.2 \pm 0.3$ and ${\rm Im}(\tau)= 0.03 \pm 0.03$. The $\chi^2$ are about the
same for the two fits. This is also consistent with pure CNI, but the
errors are even larger. These are compared in Fig.6. (Akchurin, Buttimore
and Penzo
\cite{ABP} have fit this data and some other data over a wider $t$-range
with a somewhat more complex form. They also find a large error in the
determination of $\tau$.)
\begin{figure}
\centerline{\epsfbox{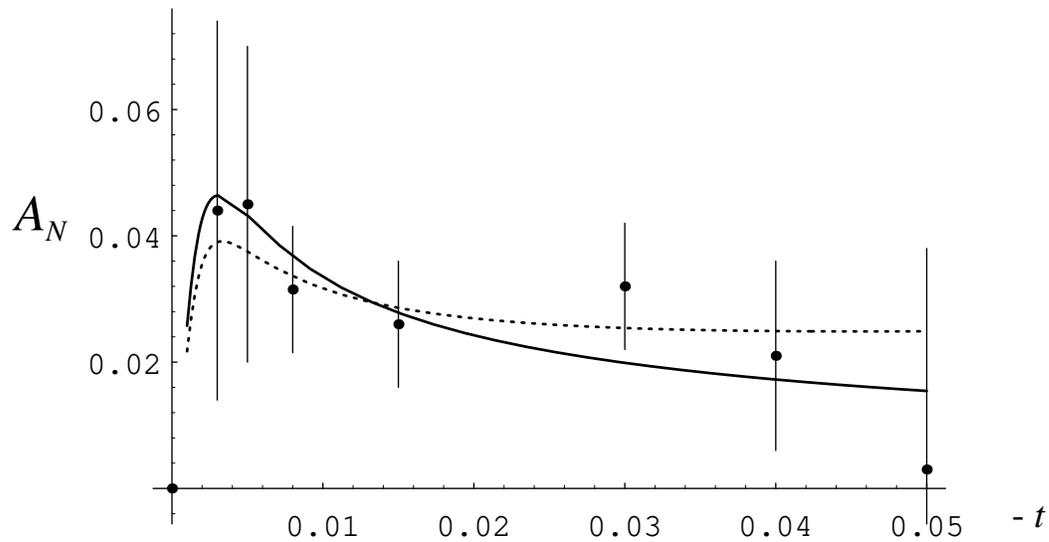}}
\medskip
\caption{Fit to E704  data with $\tau=0$ (solid) and $\tau = 0.2 + 0.03I$
(dotted)}
\end{figure}

There is an ancient ``theoretical prejudice''  \cite{AM} that
scattering amplitudes will become spin-independent at high energy.
I don't know where that comes from; there were extensive
theoretical and  phenomenological studies in the 1970's, mainly in
the context of Regge theory; for some examples, see \cite{P&K,Berger}.
These indicate no deep reason for this simplicity to occur. If
there were to be one it would have to lie within the details of
the strong interactions, of QCD applied at small $t$. After all, the
QED spin dependence does not vanish asymptotically. There are many
indications from these and more recent theoretical work that
quantitatively the spin-flip amplitude becomes small; the fit to
the energy dependence I showed demonstrates this, too. But how small?
Small enough that we can safely conclude that
$\tau$ is small enough to allow a 5\% CNI measurement of the
polarization? I believe this remains an open question. It is, I
think, an interesting question, too. Thus the two most recent
works which address the spin-flip amplitude, by Anselmino and Forte
\cite{AF} and by Goloskokov and Selyugin \cite{GS} use quite
different non-perturbative approaches to calculate this. The
former uses an instanton generated vertex; it behaves rather like a
mass insertion and leads to a small value of $\tau$ which falls
with energy as $1/\sqrt{s}$. The latter uses a
semi-phenomenolgical picture, rather like Pumplin and Kane
\cite{P&K} used years ago -- effectively Regge cuts. In this case
$\tau$ is also small but it does not fall with energy. Rather than
use these predictions as support for CNI, it seems to me important
to study them experimentally, since they represent very
different physics.

I would like thank A.Krisch, N. Buttimore, G.Bunce and Y.Makdisi for
valuable discussions.

\end{document}